\documentclass[prb,twocolumn,superscriptaddress]{revtex4}
\usepackage{amsmath,amssymb}
\usepackage{graphicx}
\newcommand{\be}{\begin{equation}}
\newcommand{\ee}{\end{equation}}

\begin{document}

\title{Invariant form of spin-transfer switching condition}

\author{Inti Sodemann}
\affiliation{Department of Physics and Astronomy, University of
South Carolina, Columbia, SC 29208, USA}

\author{Ya.\ B. Bazaliy}
\thanks{sodemann@physics.sc.edu, yar@physics.sc.edu}
 \affiliation{Department of Physics and Astronomy, University of
South Carolina, Columbia, SC 29208, USA}
 \affiliation{Institute of Magnetism, National Academy of Science,
Kyiv 03142, Ukraine}

\date{\today}

 \begin{abstract}
We derive an invariant form of the current-induced switching
condition in spin-transfer devices and show that for energy minima
and maxima the ``switching ability'' of the current is determined by
the spin torque divergence. In contrast, energy saddle points are
normally stabilized by current-induced merging with other
equilibria. Our approach provides new predictions for several
experimental setups and shows the limitations of some frequently
used approximations.

\end{abstract}


\maketitle

\begin{figure}[b]
\begin{center}
\includegraphics[scale = 0.65]{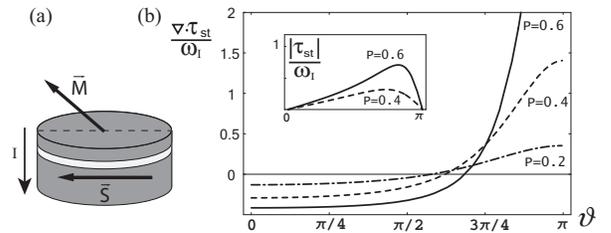}
\end{center}
\caption{(a) Typical spin-transfer device. (b) Angular dependence of
the divergence and magnitude (inset) of the Slonczewski's spin
torque term~\cite{slonczewski} for different polarizations $P$.}
\label{fig:div_and_theta*}
\end{figure}

High density electric currents induce magnetization motion and
switching in nano-size metallic wires containing alternating
ferromagnetic and non-magnetic layers
(Fig.~\ref{fig:div_and_theta*}(a)). This phenomenon is finding
important applications in computer memory and logic devices.
Switching is caused by the spin-transfer torque
$\boldsymbol{\tau}_{st}$ \cite{berger,slonczewski} which depends on
the current, spin polarization, material parameters and the geometry
of the device. Once $\boldsymbol{\tau}_{st}$ is found, the
magnetization dynamics can be obtained from the
Landau-Lifshitz-Gilbert (LLG) equation. A simple but often
sufficiently accurate approximation is the macrospin model that
assumes uniform magnetization of the layer ${\bf M}(r,t) = M {\bf
n}(t)$, where $M$ is the saturation magnetization value and ${\bf
n}$ is a unit vector. In this case the LLG equation reads
 \begin{equation}
 \label{eq:vector_LLG}
{\dot {\bf n}} = \left[ - \frac{\partial \varepsilon}{\partial {\bf
n}} \times {\bf n} \right] + \boldsymbol{\tau}_{st}({\bf n}) +
\alpha [{\bf n} \times \dot {\bf n}]
 \ ,
\end{equation}
where $\varepsilon({\bf n}) = (\gamma/M) E({\bf n})$, $E({\bf n})$
is the magnetic energy  per unit volume, $\gamma$ is the
gyromagnetic ratio, and $\alpha$ is the Gilbert damping constant.
The spin-transfer torque is proportional to electric current,
$\boldsymbol{\tau}_{st} \sim I$. At $I = 0$ vector ${\bf n}$ assumes
an equilibrium position ${\bf n}_{eq}$ at a minimum of magnetic
energy. A nonzero current has two effects: First, the spin torque
gradually shifts the equilibrium away from its original position
${\bf n}_{eq}(I = 0) \to {\bf n}_{eq}(I)$. Second, a stable
equilibrium may abruptly turn unstable at a critical current $I_c$,
causing magnetic switching.\cite{slonczewski}

Computation of $I_c$ for a given equilibrium is a straightforward
though cumbersome mathematical procedure. It would be much
simplified if one could find a single quantity that determines the
``switching ability'' of spin torque at a given equilibrium ${\bf
n}_{eq}$. Often, it is implicitly assumed that the magnitude
$|\boldsymbol{\tau}_{st}({\bf n}_{eq})|$ itself is the relevant
quantity, and, in particular, a sharp difference should exist
between the destabilization of ``collinear"
($\boldsymbol{\tau}_{st}({\bf n}_{eq}) = 0$) and ``non-collinear"
($\boldsymbol{\tau}_{st}({\bf n}_{eq}) \neq 0$) equilibria. This
view is certainly oversimplified, and it was argued
\cite{ralph_stiles_2008} that the critical current should also
depend on the derivatives of $\boldsymbol{\tau}_{st}$. Here we show
that the ``switching ability'' can be indeed introduced for extremum
(minimum or maximum) energy points and in some cases for energy
saddle points, and find explicit expressions for it. We apply our
approach to a number of experimental devices, obtain a new
qualitative understanding of their dynamics, and clarify the
limitations of some approximations.

The LLG equation (\ref{eq:vector_LLG}) can be equivalently written
as
  \be\label{sLLG}
(1+\alpha^2) \
\mathbf{\dot{n}}=\mathbf{F}(\mathbf{n})\equiv\boldsymbol{\tau}(\mathbf{n})+\alpha
\mathbf{n}\times \boldsymbol{\tau}(\mathbf{n}) \ ,
  \ee
with $\boldsymbol{\tau} = \boldsymbol{\tau}_c +
\boldsymbol{\tau}_{st}$, $\boldsymbol{\tau}_c = -
[(\partial\varepsilon/\partial {\bf n}) \times {\bf n}]$. The
equilibrium magnetization orientations ${\bf n}_{eq}$ satisfy
$\boldsymbol{\tau}({\bf n}_{eq})=0$. Their stability can be
investigated by linearizing the equation of motion. In spherical
coordinates $(\phi,\theta)$ one decomposes ${\bf F} = F^{\phi} {\bf
e}_{\phi} + F^{\theta} {\bf e}_{\theta}$ in terms of the unit
vectors ${\bf e}_{\phi}$, ${\bf e}_{\theta}$ along the coordinate
lines and obtains:
  \be
 \begin{pmatrix}
\dot{\delta\phi} \\
\dot{\delta\theta}
\end{pmatrix}
    =
 \begin{pmatrix}
\frac{1}{\sin\theta} \frac{\partial F^\phi}{\partial \phi}
    & \frac{1}{\sin\theta} \frac{\partial F^\phi}{\partial \theta} \\
\frac{\partial F^\theta}{\partial \phi}
    & \frac{\partial F^\theta}{\partial \theta}
\end{pmatrix}
 \begin{pmatrix}
\delta \phi \\
\delta \theta
\end{pmatrix}
 = \hat D  \begin{pmatrix}
\delta \phi \\
\delta \theta
\end{pmatrix}.
  \ee
Stability of an equilibrium requires both eigenvalues of the
``dynamic matrix'' $\hat D$ to have negative real parts. This is
equivalent to
  \be\label{trdet}
  {\rm Tr} \hat{D}(\mathbf{n}_{eq})<0
,
 \quad
\det \hat{D}(\mathbf{n}_{eq})>0.
  \ee
Noticing that matrix $\hat D$ is not covariant, we are led to
introduce the related matrix of covariant derivatives
$$
\hat D_{cov} =
  \begin{pmatrix}
    \frac{1}{\sin\theta} \frac{\partial F^\phi}{\partial \phi} +
    \frac{\cos\theta}{\sin\theta}F^{\theta}
    &
    \frac{1}{\sin\theta} \frac{\partial F^\phi}{\partial \theta}
    \\
    \frac{\partial F^\theta}{\partial \phi} - \cos\theta F^{\phi}
    &
    \frac{\partial F^\theta}{\partial \theta}
  \end{pmatrix} \ .
$$
The trace of $\hat D_{cov}$ is an invariant quantity equal to
$$
{\rm Tr} \hat D_{cov} = {\rm div} \mathbf{F} =
\frac{1}{\sin\theta}\left(\frac{\partial}{\partial
\theta}(\sin\theta F^\theta)+\frac{\partial F^\phi}{\partial
\phi}\right).
$$
Crucially, $\hat D = \hat D_{cov}$ at equilibrium points. Hence, an
invariant condition ${\rm div}{\bf F}<0$ can be used instead of the
first inequality in (\ref{trdet}). The latter is relevant for
treating the equilibria corresponding to the energy extrema (minima
and maxima). In this case $\det\hat D_{I = 0} > 0$, with ${\rm
Tr}\hat D_{I = 0}<0$ at the minimum and ${\rm Tr}\hat D_{I = 0}>0$
at the maximum points. Thus the condition of destabilization or
stabilization is the change of sign of the trace, i.e., of ${\rm
div}{\bf F}$. Using the relation between $\mathbf{F}$ and
$\boldsymbol{\tau}$ and notation ${\rm div} \mathbf{F} =
\nabla\cdot\mathbf{F}$, where $\nabla$ operates in the $\bf
n$-space, we find:
 \be\label{divF1}
 \nabla\cdot\mathbf{F} = \nabla\cdot\boldsymbol{\tau}
 -\alpha [\nabla \times \boldsymbol{\tau}] \cdot {\bf n},
 \ee
where $[\nabla\times\boldsymbol{\tau}]  \cdot {\bf n} =
 - (\partial \tau_\theta / \partial
\phi -  \partial(\sin \theta\tau_\phi) / \partial \theta) /\sin
\theta$. The general expression for the spin-torque created by a
polarizer pointing along the unit vector $\bf s$ reads:
 \be\label{vectau}
 \boldsymbol{\tau}_{st}({\bf n},I) = \omega_I \ g(\mathbf{n} \cdot {\bf s})  \
 [\mathbf{n}\times(\mathbf{s}\times\mathbf{n})]
 \equiv \omega_I {\bf f}_{st}({\bf n})
 \ ,
 \ee
where $\omega_I =  (\gamma/M) (\hbar I/2eV)$, $V$ is the magnetic
layer volume, $e$ is the electron charge, and $g(\mathbf{n} \cdot
{\bf s})$ is the efficiency factor.\cite{bjz2004} Using
$\nabla\cdot\boldsymbol{\tau}_c = 0$ and $[\nabla \times
\boldsymbol{\tau}_{st}] \cdot {\bf n} = 0$ we get
 \be\label{diverg}
\nabla\cdot\mathbf{F} = \nabla \cdot \boldsymbol{\tau}_{st} - \alpha
\nabla \times \boldsymbol{\tau}_c = \omega_I \ \nabla \cdot {\bf
f}_{st}-\alpha  \nabla^2\varepsilon \ .
 \ee
Here the first term is proportional to the current and can lead to
the sign change of the whole expression. We see that the switching
ability is determined by the divergence $\nabla\cdot {\bf f}_{st}$
that characterizes the angular dependence of the spin torque. Note
that with Eq.~(\ref{diverg}) condition ${\rm div}{\bf F} = 0$ can be
viewed as a limiting case of the condition for the existence of a
precession cycle (P), $\oint (\boldsymbol{\tau}_{st} \cdot {\bf
e}_{\perp}) dl = \alpha\oint ([\nabla\varepsilon \times {\bf
n}]\cdot{\bf e}_{\perp} dl$, where integrals are taken along the
cycle.\cite{serpico} In terms of the bifurcation
theory,\cite{crawford1991} local destabilization of the minimum
points is the {\em Hopf bifurcation} which normally produces a
stable precession cycle.

Consider now the experimentally relevant case of small Gilbert
damping, $\alpha \ll 1$. Expression (\ref{diverg}) shows that the
critical current satisfies $I_c \propto \alpha$ and hence will also
be small. Therefore at $I = I_c$ the equilibrium point will be close
to the zero current equilibrium, ${\bf n}_{eq}(I_c) = {\bf
n}_{eq}(0) + \Delta {\bf n}$ with $\Delta {\bf n} \propto I_c
\propto \alpha$. Expanding (\ref{diverg}) up to linear terms in
$\alpha$ we get an approximate stability condition
 \be\label{approxdivF}
 \omega_I \ \nabla \cdot {\bf f}_{st} \big|_{n_{eq0}} \leq \alpha
\nabla^2\varepsilon \big|_{n_{eq0}}
 \ee
with equality achieved at the critical current. Importantly, all
quantities in (\ref{approxdivF}) are evaluated at the unperturbed
equilibrium point ${\bf n}_{eq}(0)$. In comparison, using conditions
(\ref{trdet}) one needs to perform an explicit calculation of ${\bf
n}_{eq}(I_c)$ even in the case of the first order expansion in
$\alpha$ (e.g.~Ref.~\onlinecite{bjz2004,sodemann}). This welcome
simplification stems from ${\rm div}{\bf F}|_{I=0} \sim \alpha$
holding for any $\bf n$, while ${\rm Tr}\hat D|_{I=0} \sim \alpha$
holds only at ${\bf n}_{eq}(0)$.

For $\boldsymbol{\tau}_{st}$ given by Eq.~(\ref{vectau}) one gets
 \be\label{divst}
\nabla\cdot {\bf f}_{st} =
 - \frac{1}{\sin\vartheta}
 \frac{d}{d\vartheta} \big(g(\cos\vartheta) \sin^2\vartheta\big).
 \ee
where $\vartheta$ is the angle between $\bf s$ and $\bf n$.
Representative graphs of $\nabla\cdot{\bf f}_{st}(\vartheta)$ are
shown in Fig.\ref{fig:div_and_theta*}(b) for the Slonczewski form
\cite{slonczewski} of $g(\vartheta)$. We observe that: (a)
Divergence $\nabla \cdot \boldsymbol{\tau}_{st}$ can substantially
differ from $|\boldsymbol{\tau}_{st}|$, i.e., the destabilization of
noncollinear equilibria may actually require larger current. (b) The
switching ability $\nabla~\cdot~{\bf f}_{st}$ vanishes at a critical
angle $\vartheta_*$ (Fig.\ref{fig:div_and_theta*}(b)).
Equation~(\ref{approxdivF}) predicts infinite critical current for
the equilibrium points lying on the ``critical circle'' (CC) defined
by $\vartheta(\phi,\theta) = \vartheta_*$  (more precisely, at CC
the approximation (\ref{approxdivF}) breaks down and $I_c$ is just
large). The critical circle divides the unit sphere into two parts.
Spin-transfer torque destabilizes the energy extrema in one of them
(which one -- depends on the current direction), while in the other
it makes them more stable. (c) The signs of $I_c$ are opposite for
equilibria located on different sides of a CC. This circumstance is
especially relevant when one considers different models of
$g(\vartheta)$. For example in the Slonczewski's case $\vartheta_*$
depends on the spin polarization $P$ and varies from
$\vartheta_{*(P=0)}\gtrsim \pi/2$ to $\vartheta_{*(P=1)} = \pi$. In
contrast, for a popular approximation $g = {\rm const}$, one has
$\vartheta_* = \pi/2$ independently of $P$. The difference between
the models becomes crucial for an equilibrium located between the
respective CC's: a given current would have a stabilizing effect in
one model, and destabilizing in another.

\begin{figure}[b]
\vspace{-0.2cm}
\includegraphics[scale=0.4]{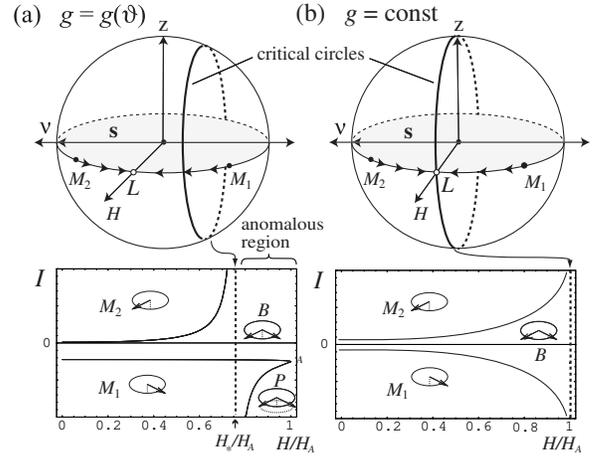}
\caption{Critical circles and the ``anomalous'' stabilization
region. Upper panels: collinear device with ${\bf s} ||
\boldsymbol{\nu}$ and $H \perp \hat z$, $H \perp \boldsymbol{\nu}$.
The energy minimum points $M_1$ and $M_2$ move with increasing $H$
as shown by the arrows. Critical circles shown for (a) generic
$g(\vartheta)$ and (b) $g = const$. Lower panels: switching
diagrams. In the regions $M_1$ and $M_2$ one equilibrium is stable,
in $B$ both are stable, and in $P$ both are unstable.}
\label{fig:collinear_device}
\end{figure}

As an illustration, consider a typical nanopillar \cite{katine} with
an $(x,y)$ easy plane and an easy axis $\boldsymbol{\nu} \perp \hat
z$, so that $\varepsilon = \frac{1}{2}\omega_p ({\bf n} \hat z)^2 -
\frac{1}{2}\omega_a ({\bf n} \boldsymbol\nu)^2 - \gamma ({\bf
H}\cdot{\bf n})$. Magnetic field is in-plane perpendicular, ${\bf H}
\perp \hat z$, ${\bf H} \perp \boldsymbol\nu$, and the polarizer
direction is $\bf s || \boldsymbol\nu$
(Fig.~\ref{fig:collinear_device}). In this setup the energy minima
$M_1$ and $M_2$ are located at $\pm \boldsymbol\nu$ at $H = 0$, move
towards each other with increasing $H$, and finally merge with the
saddle point $L$ as the field reaches the easy axis anisotropy field
$H_A = \omega_a/\gamma$. The switching diagrams for a generic
$g(\vartheta)$ dependence (Fig.~\ref{fig:collinear_device}a) and the
special case of $g = {\rm const}$ (Fig.~\ref{fig:collinear_device}b)
are qualitatively different \cite{smith,sodemann} with the former
displaying the ``anomalous'' region
(Fig.~\ref{fig:collinear_device}a). In our approach the ``anomaly''
is naturally explained by the fact that the minimum point $M_1(H)$
crosses the critical circle at $H = H_*$. For $g = const$ the minima
never cross the critical circle, hence the anomalous region is
absent. Note that the $\tau_{st}(\vartheta)$ dependence produced by
the $g = {\rm const}$ approximation is qualitatively similar to the
actual one. Nevertheless, it does not lead to the correct
qualitative picture of switching when the equilibria of interest are
close to the actual critical circle.

A similar example is provided by the nanopillars with ${\bf H} ||
\boldsymbol{\nu}$ where magnetic field causes a crossing of the
critical circle by an energy maximum point. That crossing naturally
explains the peculiar sign change of the corresponding critical
current found in Ref.~\onlinecite{bjz2004}.

\begin{figure}[b]
\vspace{-0.2cm}
\includegraphics[scale=0.4]{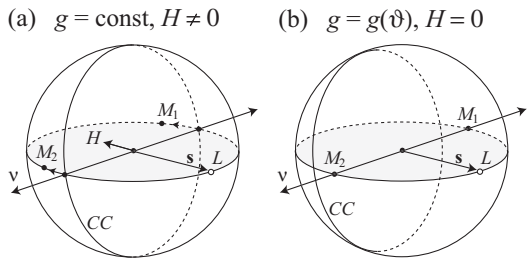}
\caption{Spin-flip transistor geometry. (a) For $g = const$ local
destabilization of the energy minima requires external field $H$ to
shift $M_{1,2}$ away from the critical circle $CC$. (b) For $g =
g(\vartheta)$ the minima are away from CC even at $H = 0$.}
 \label{fig:sft_precession}
\end{figure}

Sensitivity to the $g(\theta)$ angular dependence turns out to be of
crucial importance for the interpretation of the ``spin-flip
transistor'' (a nanopillar with ${\bf s} \perp \hat z$, ${\bf s}
\perp \boldsymbol\nu$) precession experiment.\cite{devolder2007}
Here the calculations with $g = const$ \cite{wang,morise} forbid
precession cycles (P) at zero magnetic field, but find them in
external field $H$ antiparallel to $\bf s$. Based on this,
Ref.~\onlinecite{devolder2007} interpreted the observation of
precession at $H = 0$ as an indication that an additional
``field-like" term had to be introduced in Eq.~(\ref{vectau}).
Within the framework of our analysis, the absence of P states at
zero field is due to the fact that at $g = const$ the $M$-points
stay on the critical circle (Fig.~\ref{fig:sft_precession}a) and
cannot be destabilized. The antiparallel field is required to shift
the $M$-points away from CC. However, for general $g(\vartheta)$,
the $M$-points are away from CC even at $H = 0$
(Fig.~\ref{fig:sft_precession}b). They can be locally destabilized,
producing P states by Hopf bifurcation without any field-like
terms.\cite{numeric} The $H = 0$ results of Refs.~\onlinecite{wang}
and \onlinecite{morise} are sensitive to the angular dependence
$g(\vartheta)$ in a manner that would be hard to foresee without the
notion of a critical circle.

\begin{figure}[t]
 \center
\includegraphics[scale=0.4]{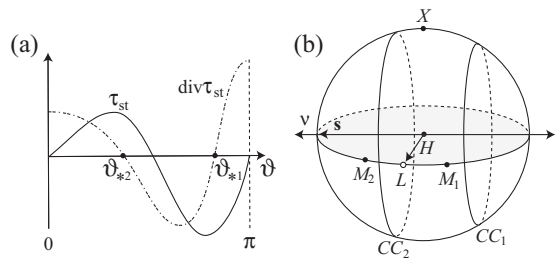}
\caption{(a) ``Wavy'' $\tau_{st}(\vartheta)$ dependence (solid line,
Ref. \onlinecite{non-monotonic}) and corresponding
$\nabla\cdot\tau_{st}$ (dashed line). (b) Critical circles
$CC_{1,2}$ and positions of minimum points $M_{1,2}$ at an
intermediate value of field $H$. The energy maximum point is $X$.}
 \label{fig:wavy}
\end{figure}

To further demonstrate the power of the analysis based on
Eqs.~(\ref{approxdivF}) and (\ref{divst}), consider experiment
\cite{non-monotonic} performed on a nanopillar device with an
unusual ``wavy'' $\tau_{st}(\vartheta)$ dependence
(Fig.~\ref{fig:wavy}a). In this case there are two critical circles,
$CC_{1}$ and $CC_{2}$, defined by the angles $\vartheta_{*1,2}$. At
zero external field the energy minima $M_{1,2}$ fall into the
regions of the same sign of $\nabla\cdot\boldmath\tau_{st}$ and can
be destabilized simultaneously, producing a precession
cycle.\cite{non-monotonic} With increasing current, the cycle
gradually approaches the energy maximum point $X$. Eventually
spin-transfer stabilizes that point \cite{bjz2004} by closing the
contour on it.\cite{non-monotonic} The notion of critical circles
suggests an experiment capable of providing additional evidence for
the ``wavy'' $\tau_{st}(\vartheta)$ dependence. If an in-plane
perpendicular $H$ is applied (Fig.~\ref{fig:wavy}b), the energy
minima $M_{1,2}$ are shifted towards the saddle point~$L$. As
$\vartheta_{*1,2}$ are not symmetric w.r.t. $\pi/2$, there will be
an interval of fields where $M_1$ had already crossed $CC_1$ and
moved into the middle region, while $M_2$ remains in the left
region. In this interval $\nabla\cdot\boldmath\tau_{st}$ has
opposite signs for $M_{1,2}$ and normal switching between $M_1$ and
$M_2$ will be possible. Further increase of $H$ will put both
$M$-points into the middle region, where they will be again
destabilized by the same current direction However, in contrast with
the $H=0$ case, now the same current direction will also destabilize
$X$, so the P state evolution will be different.\cite{numeric}

\begin{figure}[b]
\center
\includegraphics[scale=0.45]{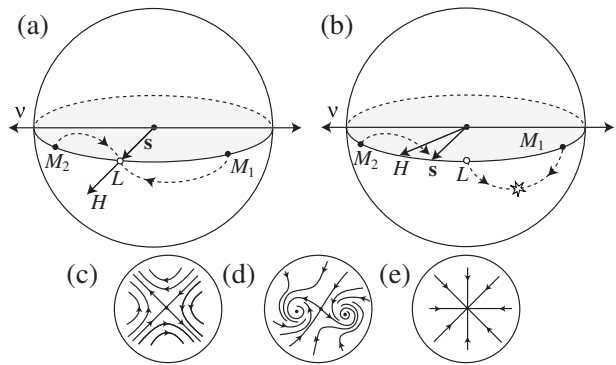}
\caption{(a) Spin-flip transistor:  ${\bf s}\perp \hat z$, ${\bf
s}\perp \boldsymbol{\nu}$, $\bf H || {\bf s}$. The dashed lines show
how the positions of the equilibria $M_1$ and $M_2$ change with
increasing current and merge with the saddle $L$. (b) General
in-plane directions of ${\bf s}$ and $\bf H$. The saddle merges with
one of the minima, while the other one asymptotically approaches
$\bf s$. (c-e) transformation of the field $\bf F$ during the
merging of a saddle with two foci in case (a).}
 \label{fig:MvsI}
\end{figure}

Let us now turn to stabilization of the saddle points. Here ${\rm
det}\hat D_{I = 0} < 0$, so the process requires a change of sign of
${\rm det}\hat D$. An example is provided by a spin-flip transistor
where the spin torque attracts $\bf n$ to the saddle point and
eventually stabilizes $L$.\cite{wang,morise} Notably, stabilization
is always accompanied by a simultaneous discontinuous change in the
nature of other equilibria. At $H = 0$ the $M$-points loose their
stability just as $L$ becomes stable.\cite{wang} With the field
parallel to $\bf s$ (note the difference with the antiparallel case
discussed above) the current leads to a significant deviation of the
$M$-points from their initial positions (Fig.~\ref{fig:MvsI}(a)). At
the critical current, $M_{1,2}$ approach $L$ and merge with it,
forming a stable center. We start by explaining why those
simultaneous transformations are not a coincidence. The saddle point
is stabilized by becoming a stable center. As topological defects of
the vector field $\bf F$, saddles and centers differ in the winding
number \cite{mermin1979} which is a topological characteristic equal
to $n = -1$ for a saddle and $n = 1$ for a center or focus. Since
the total winding number is conserved (the Poincar\'e index
theorem), a saddle point cannot be transformed into a center
locally. The saddle-to-center transformation has to either proceed
via merging with other defects (Fig.~\ref{fig:MvsI}(c-e)), or be
accompanied by a simultaneous change of nature of the far away
equilibria.

\begin{figure}[t]
\includegraphics[scale=0.4]{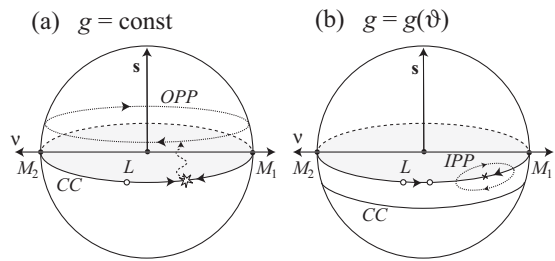}
\caption{``Magnetic fan'' geometry. (a) At $g = const$ the point $M$
stays on $CC$ until it collides with $L$, creating a large ``OPP
cycle'' (dotted line). (b) At $g = g(\vartheta)$ $M$ is away from
$CC$. Its local destabilization can create a small ``IPP cycle''.}
 \label{fig:magnetic_fan}
\end{figure}

More insight comes from considering a generic case of ${\bf H}$ and
${\bf s}$ pointing in arbitrary in-plane directions
(Fig.~\ref{fig:MvsI}(b)). Here $L$ merges with one of the minima
annihilating both equilibria, while the other minimum approaches
$\bf s$. Such merging is allowed by the winding number conservation
and in fact the bifurcation theory \cite{crawford1991} shows that it
is the most general case of the {\em saddle-node bifurcation}; i.e.,
the saddle point is normally not stabilized but rather destroyed in
a collision with an energy extremum point. Stabilization happens
only in special circumstances, such as ${\bf s}$ pointing exactly
into $L$. In this case $L$ remains an equilibrium for the
arbitrarily large  current and cannot disappear. That restriction
produces the {\em transcritical bifurcation} \cite{crawford1991}
where the energy extremum and the saddle exchange their nature in a
collision. We find the critical current to be
 \be \label{eq:s||L}
\omega_I = \frac{
    \sqrt{-{\rm det}\hat D|_{I = 0,\alpha = 0}}
    }{g(0)} \ .
 \ee
The above formula remains a good estimate in the case of a small
misalignment between ${\bf s}$ and $L$. It also works for the spin
flip transistor (Fig.~\ref{fig:MvsI}a) even though here additional
symmetries produce a more rare {\em fork bifurcation}.

To sum up, the saddle point stabilization is associated with a
merging or a close approach of equilibria, and thus can be detected
without even calculating the dynamic matrix. A non-local bifurcation
can only occur in devices of exceptionally high symmetry.

Local destabilization and equilibrium merging are two alternative
switching mechanisms which can compete with each other. Consider the
${\bf s} || \hat z$ ``magnetic fan'' experiment
\cite{houssameddine2007}(Fig.~\ref{fig:magnetic_fan}). Here the $g =
const$ approximation is special since the critical circle goes
through the $M$-points. It predicts a merging of $L$ and $M_1$,
after which the system jumps into an ``OPP cycle''
(Fig.~\ref{fig:magnetic_fan}a). For angle-dependent $g(\vartheta)$
the $CC$ is away from the $M$-points, allowing for a competing
scenario with local destabilization of $M_1$ producing an ``IPP
cycle'' (Fig.~\ref{fig:magnetic_fan}b).\cite{numeric} We estimate
the critical currents as $\omega_{I(IPP)} \approx
\alpha\omega_p/g'(\pi/2)$ and $\omega_{I(OPP)} \approx
\omega_a/g(\pi/2)$. Using Slonczewski's $g$ and experimental
parameters \cite{houssameddine2007} we find
$\omega_{I(OPP)}/\omega_{I(IPP)} \approx 0.2$, which yields an OPP
cycle scenario in accord with experiment. Furthermore, we predict
that a device with a sufficiently large $\omega_a$ would manifest an
IPP cycle.

To conclude, we found that the ability of electric current to
destabilize magnetization at its energy minimum (or to force it to
be stable at its energy maximum) is determined by the spin torque
divergence. We also showed that the sphere swept by the
magnetization vector is divided into stabilization and
destabilization regions by critical circles. Finally, we
demonstrated that saddle points are stabilized via a topologically
distinct route of merging with other equilibria, and discussed the
competition between such a merging and local destabilization.

It is our pleasure to thank S. Garzon, R. A. Webb, and R. R.
Ramazashvili for stimulating discussions.


\begin{thebibliography}{}
\bibitem{numeric} We have confirmed these qualitative
predictins  numerically.
\end{thebibliography}
\end{document}